\documentclass{manuscript}
\usepackage[T1]{fontenc}
\usepackage{times}

\begin{document}

\title{A low-communication-overhead parallel method for the 3D incompressible Navier-Stokes equations}

\author{\normalsize 
        Jiabin Xie$^{1}$, \ 
        Jianchao He$^{1}$, \ 
        Yun Bao$^{1,2}$\footnotemark[1]~, \ 
        Xi Chen$^{2}$\footnotemark[1] \\
        \normalsize
        $^{1}${School of Aeronautics and Astronautics, Sun Yat-sen University, Guangzhou, China.} \\
        \normalsize
        $^{2}${Institute of Fluid Mechanics, Beihang University, Beijing, China.}
}
\renewcommand{\thefootnote}{\fnsymbol{footnote}}
\footnotetext[1]{
    Corresponding author.
    \par{Yun Bao, E-mail: stsby@mail.sysu.edu.cn}
    \par{Xi Chen, E-mail: chenxi97@outlook.com}}

\date{}
\maketitle

\addcontentsline{toc}{section}{Abstract}
\begin{abstract}
  This paper presents a low-communication-overhead parallel method for solving the 3D incompressible Navier-Stokes equations. 
  A fully-explicit projection method with second-order space-time accuracy is adopted. 
  Combined with fast Fourier transforms, the parallel diagonal dominant (PDD) algorithm for the tridiagonal system is employed 
  to solve the pressure Poisson equation, differing from its recent applications to compact scheme derivatives computation 
  (Abide et al. 2017\cite{Abide2017}) and alternating-direction-implicit method (Moon et al. 2020\cite{Moon2020}). 
  The number of all-to-all communications is decreased to only two, in a 2D pencil-like domain decomposition. 
  The resulting MPI/OpenMP hybrid parallel code shows excellent strong scalability up to $10^4$ cores and small wall-clock time per timestep. 
  Numerical simulations of turbulent channel flow at different friction Reynolds numbers ($Re_{\tau}$ = 550, 1000, 2000) have been conducted and the 
  statistics are in good agreement with the reference data. 
  The proposed method allows massively simulation of wall turbulence at high Reynolds numbers as well as many other incompressible flows.
\end{abstract}


\section{Introduction}\label{sec1}

Understanding wall turbulence is of great importance for both fundamental physics and engineering applications. Owing to the 
progress of computing technology and the development of numerical methods, large-scale direct numerical simulation (DNS) of 
wall turbulence has become feasible. Presently, the numerical methods for DNS of wall turbulence are either spectral method 
\cite{LM2015,MKM1999,LJ2014} or finite difference method\cite{BPO2014}. The spectral method is only suitable for flow simulation with simple boundary conditions, 
but with high data accuracy hence appreciated much by academic study. In contrast, the finite difference method greatly eases 
restrictions on boundary conditions and enables the simulation of complex flows\cite{Verzicco2000} provided with the immersed boundary method. 
Though not as accurate as spectral code, it has been shown that adequate first- and second-order statistics can be obtained by 
lower-order finite difference schemes when grid resolution is sufficiently high\cite{Vreman2014}. However, both methods suffer from the issue 
that a huge demand for memory and computing power is required which limits the application of DNS for turbulent flows at 
practically high Reynolds number. Taking a simple geometry wall-flow, i.e. turbulent channel, for example, current available DNS 
data are mostly in the friction-Reynolds-number range $Re_{\tau}$ = 180 -- 2000\cite{MKM1999,KMM1987,HJ2006}, whilst data of higher Reynolds Numbers 
($Re_{\tau}$ = 4000, 4200, 5200, and 8000)\cite{BPO2014,LJ2014,LM2015,Yamomoto2018}, though reported, remain relatively fewer. 

The most time-consuming part of a numerical simulation of incompressible flow is to solve the pressure Poisson equation. Among 
the solving methods (e.g., the multigrid method, the fast multipole method), Gholami et al.\cite{Gholami2016} show that the fast Fourier 
Transform (FFT) based direct method is more efficient and accurate. The reason is that, using the FFT technique, the pressure Poisson 
equation can be decoupled into a series of tridiagonal linear equations, which can be solved by the Thomas algorithm hence saving 
a lot of computation time. The FFT method has already been applied to numerical simulation of incompressible flows\cite{Schumann1988}, and recently, 
applied to large-scale parallel numerical simulation of several turbulent flows such as Rayleigh-B\'enard convection and 
Tayler-Couette flow\cite{vanderPoel2015,Zhu2018,Ostilla2016,Costa2018}.

Note that there are two challenges to implement the large-scale parallelization of the FFT-based direct method, i.e., to improve the 
parallel efficiency of the multi-dimensional FFT calculation in the multi-dimensional decomposition domain, and to exploit the 
parallelization potential efficiently for solving tridiagonal systems. To achieve the formal goal, open source libraries for parallel 
multi-dimensional FFT (e.g., PFFT\cite{PFFT}, P3DFFT\cite{P3DFFT} and 2DECOMP\&FFT\cite{LiLaizet2010}) can be used which typically adopts 2D pencil-like domain 
decomposition for better scalability. Although it unavoidably introduces multiple all-to-all global transpose communications, several 
optimization methods (e.g. minimizing communication amount and computation-communication overlap) can be applied jointly to ensure the 
parallel efficiency\cite{Duy2014,Song2016}. However, for the latter goal of solving tridiagonal systems which is implicit in nature, it brings in a 
challenge to the parallel implementation of FFT-based direct method in a multi-dimensional domain decomposition. The existing practice 
\cite{vanderPoel2015,Costa2018} is to keep the computational domain undecomposed in the implicitly coupling direction by all-to-all data transposes, so that 
the Thomas algorithm can be used. There are several parallel algorithms to solve tridiagonal systems, including the cyclic doubling 
method\cite{Stone1973}, the cyclic reduction method\cite{Hockney1965} and a class of methods based on the idea of divide and conquer\cite{Lawrie1984,Wang1981,Sun1989}. Among them, the 
parallel diagonal dominant (PDD) algorithm proposed by Sun et al.\cite{Sun1989} is of our interest as it is featured by low communication 
overhead and high parallel efficiency. When the linear system is diagonal dominant, it can provide an approximate solution with error 
smaller than machine accuracy.

Although the PDD algorithm has been proposed since the 1990s, its applications in computational fluid dynamics are still limited. 
Abide et al. (2017)\cite{Abide2017} employed the PDD algorithm to address the computations of compact derivatives and interpolations. 
It demonstrates that the PDD algorithm not only maintains the accuracy and conservation features but also contributes to a 
good parallel performance up to 4096 cores.
Bao et al. (2017)\cite{Bao2017} introduced the PDD algorithm to solve the pressure Poisson equation for the 2D Rayleigh-B\'enard 
convection. 
Recently, Moon et al. (2020)\cite{Moon2020} proved the applicability of the PDD algorithm for non-diagonal dominant systems derived 
from discretization of incompressible momentum equations. The accuracy of results is found to be determined by the CFL number and 
the number of grid points in each decomposed block.

This paper aims to develop a highly efficient parallel method for solving the 3D incompressible Navier-Stokes equations and use it to 
accomplish wall turbulence DNS at high Reynolds numbers. Our approach is based on the projection method with the second-order space-time 
accuracy, and the main idea is to establish an approximate solver for the pressure Poisson equation, which combines 2D FFTs with the PDD 
algorithm for the tridiagonal system. Using MPI/OpenMP hybrid model and 2D pencil-like domain decomposition, the resulting code, named 
by us as PowerLLEL (a Powerful paraLLEL solver for incompressible flows), has a very good parallel performance (demonstrated below). 
We believe that not only wall turbulence but also other incompressible flows can be simulated efficiently by this method.

This paper is organized as follows. Section \ref{sec2} introduces the numerical method adopted in this paper for channel flow. Section \ref{sec3} details 
the parallel strategy and the efficient implementation of the numerical method for massively parallel simulation. Section \ref{sec4} presents 
the computational performance of our code. In Section \ref{sec5}, several DNS cases of channel flow are presented to validate our code. Finally, 
Section \ref{sec6} presents the main conclusion and an outlook of future works.

\section{Numerical method}\label{sec2}

\begin{figure}[ht]
  \centerline{\includegraphics[width=200pt]{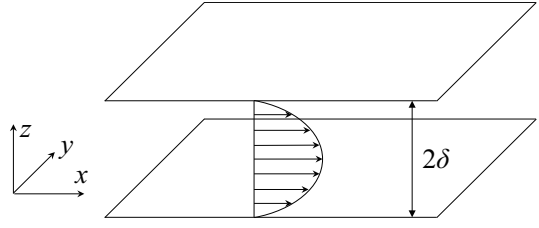}}
  \caption{Schematic of channel flow. The streamwise, spanwise and wall-normal 
  directions are denoted by $x$, $y$ and $z$, respectively. $\delta$ is the 
  channel half-height and $U$ is the streamwise mean velocity.\label{fig1}}
\end{figure}

One of the canonical wall turbulences is channel flow. As illustrated in Figure \ref{fig1}, the flow between two infinite parallel planes 
is driven by a unidirectional pressure gradient. In order to simulate the flow in an infinite domain, periodic boundary conditions 
are used in the streamwise ($x$) and spanwise ($y$) directions. A no-slip boundary condition is imposed at the two planes in the 
wall-normal direction ($z$). The governing 3D incompressible Navier-Stokes equations are as follows
\begin{align}
  \nabla \cdot \bf u &= 0, \label{eq1}\\
  \frac{\partial \mathbf{u}}{\partial t} + (\mathbf{u} \cdot \nabla)\mathbf{u} &= -\nabla p + \frac{1}{Re} \nabla^2 \mathbf{u}, \label{eq2}
\end{align}
where $\mathbf{u}$ is the velocity vector, $p$ is the pressure, $t$ is time and $Re$ is the Reynolds number.

\begin{figure}[ht]
  \centerline{\includegraphics[width=200pt]{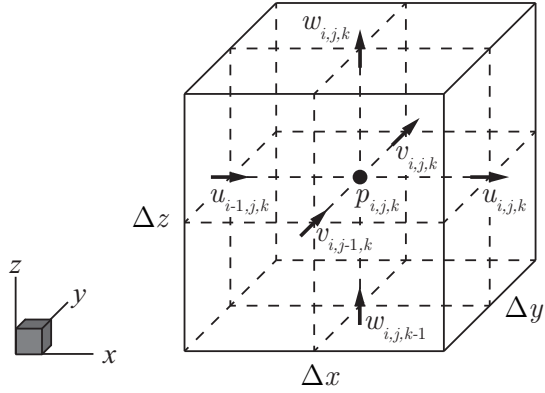}}
  \caption{Schematic of staggered grids. Velocity vectors are placed at the 
  cell surface while pressure is placed in the cell center.\label{fig2}}
\end{figure}

Considering the flow characteristics and the boundary conditions mentioned above, the Navier-Stokes equations are discretized on a 
Cartesian grid, uniformly-spaced in the streamwise and spanwise directions and nonuniformly-spaced in the wall-normal direction. 
The spatial discretization scheme is the standard second-order centered finite difference with a staggered arrangement of velocity 
and pressure variables, as illustrated in Figure \ref{fig2}. Time advancement is performed with an explicit second-order Runge-Kutta 
scheme. Therefore, the projection method used in this paper is as follows.
\begin{align}
  \mathbf{u}^* &= \mathbf{u}^n + \Delta t \left[ -N(\mathbf{u}^n) + L(\mathbf{u}^n) \right], \label{eq3} \\
  \mathbf{u}^{n+1} &= \mathbf{u}^* - \Delta t~\nabla p^{n+1}, \label{eq4}
\end{align}
where $u^*$ is the provisional velocity, $N(\mathbf{u})=(\mathbf{u} \cdot \nabla)\mathbf{u}$ and $L(\mathbf{u})=\nabla^2 \mathbf{u}/Re$ are the 
linear and nonlinear terms respectively. The second-order Runge-Kutta scheme is applied to equation (\ref{eq3}) to calculate the 
provisional velocity $u^*$.
\begin{align}
  \overline{\mathbf{u}^*} &= \mathbf{u}^n + \Delta t \left[ -N(\mathbf{u}^n) + L(\mathbf{u}^n) \right], \label{eq5} \\
  \mathbf{u}^* &= \mathbf{u}^n + \frac{1}{2} \Delta t \left[ -N(\mathbf{u}^n) + L(\mathbf{u}^n) - N(\overline{\mathbf{u}^*}) + L(\overline{\mathbf{u}^*}) \right]. \label{eq6}
\end{align}
The pressure Poisson equation is derived from the combination of the divergence operator with the velocity correction equation 
(\ref{eq4}):
\begin{equation}\label{eq7}
  \nabla^2 p^{n+1} = \frac{\nabla \cdot \mathbf{u}^*}{\Delta t}.
\end{equation}
Then $p^{n+1}$ is obtained. The final velocity at the new timestep $u^{n+1}$ can be updated according to equation (\ref{eq4}). 
The velocity boundary conditions should be imposed on the provisional/final velocity at each substep or timestep, while the 
pressure boundary conditions are imposed only after the pressure Poisson equation is solved.

\subsection{The pressure Poisson equation}

For the numerical simulation of incompressible flows, the numerical solution of the pressure Poisson equation (\ref{eq7}) 
is often difficult to tackle, because it couples all the points in the three-dimensional domain. For simplicity, consider 
the following Poisson equation discretized with second-order centered finite difference on a uniform grid:
\begin{equation}\label{eq8}
  \frac{p_{i-1,j,k} - 2 p_{i,j,k} + p_{i+1,j,k}}{\Delta x^2} +
  \frac{p_{i,j-1,k} - 2 p_{i,j,k} + p_{i,j+1,k}}{\Delta y^2} +
  \frac{p_{i,j,k-1} - 2 p_{i,j,k} + p_{i,j,k+1}}{\Delta z^2} = f_{i,j,k}.
\end{equation}
According to the boundary conditions, we apply corresponding type discrete Fourier transform (DFT) operators $\mathcal{F}$ 
to equation (\ref{eq8}) in $x$ and $y$ directions respectively, and then equation (\ref{eq8}) is decoupled to a series 
of tridiagonal linear equations
\begin{equation}\label{eq9}
  \left( \frac{\lambda_i}{\Delta x^2} + \frac{\lambda_j}{\Delta y^2} \right) \hat{\hat{p}}_{i,j,k} + 
  \frac{1}{\Delta z^2}\left( \hat{\hat{p}}_{i,j,k-1} + 2\hat{\hat{p}}_{i,j,k} +\hat{\hat{p}}_{i,j,k+1} \right) = 
  \hat{\hat{f}}_{i,j,k},
\end{equation}
where $\hat{\hat{\square}} \equiv \mathcal{F}^y (\mathcal{F}^x (\square))$, $\lambda_i$ and $\lambda_j$ are eigenvalues.
Actually, the operator $\mathcal{F}$ could be applied a third time in $z$ direction and thus equation (\ref{eq9}) could 
be completely decoupled. However, considering that a nonuniform grid in $z$ direction is common in the numerical 
simulation of wall turbulence, the Thomas algorithm should be a more flexible solution method to equation (\ref{eq9}). 
Furthermore, the time complexity of the Thomas algorithm, $O(n)$, is better than that of the DFT algorithm, $O(n\log n)$.
Hence, the Thomas algorithm is applied to solve the equation (\ref{eq9}) efficiently. After that, we apply inverse DFT 
operators $\mathcal{F}^{-1}$ to the solution in $x$ and $y$ directions respectively, and then the final numerical 
solution of the pressure Poisson equation can be obtained.
Note, however, that the Thomas algorithm imposes a restriction on the domain decomposition, i.e., the computational 
domain should not be decomposed in $z$ direction. From the discussion in Section \ref{sec3} below, two additional 
all-to-all transpose communications are required to perform the Thomas algorithm in a 2D domain decomposition, which 
degrades the parallel performance to some extent. Hence, this paper adopts the highly efficient PDD algorithm with low 
communication overhead to solve the tridiagonal systems (\ref{eq9}) in parallel.

\subsection{The PDD algorithm}

The following is a brief introduction to the PDD algorithm. Consider the following tridiagonal linear system:
\begin{equation}\label{eq10}
  Ax=d,
\end{equation}
where $A$ is a tridiagonal matrix of order $n$, $x=(x_1,\dots,x_n )^T$ and $d=(d_1,\dots,d_n )^T$. In order to solve equation 
(\ref{eq10}) in parallel, the matrix $A$ should be partitioned into $p$ sub-matrices, where $p$ is the number of partitions. 
For convenience, the order $n$ is assumed to be exactly divisible by $p$, i.e., $n=pm$, where $m$ is the size of a 
partition. The matrix $A$ can be regarded as the superposition of the following two matrices
\begin{equation}\label{eq11}
  A = \tilde{A} + \Delta A,
\end{equation}
where $\tilde{A}$ is a block diagonal matrix with diagonal sub-matrices $A_i (i=1,\dots,p)$ of order $m$, and the remaining 
elements (the entries with frame) make up $\Delta A$, as illustrated in equation (\ref{eq12}). Note that 
$\square_j^i=\square_{(i-1)*m+j}$. The vectors $x$ and $d$ are partitioned equally by $p$.
\begin{equation}\label{eq12}
  \setcounter{MaxMatrixCols}{15}
  A = 
  \begin{pmatrix}
    b_1^1 & c_1^1 & & & & & & & & & & & & & \\
    a_2^1 & b_2^1 & c_2^1 & & & & & & & & & & & & \\
     & \ddots & \ddots & \ddots & & & & & & & & & & & \\
     & & a_{m-1}^1 & b_{m-1}^1 & c_{m-1}^1 & & & & & & & & & & \\
     & & & a_m^1 & b_m^1 & \boxed{c_m^1} & & & & & & & & & \\
     & & & & \boxed{a_1^2} & b_1^2 & c_1^2 & & & & & & & & \\
     & & & & & a_2^2 & b_2^2 & c_2^2 & & & & & & & \\
     & & & & & & \ddots & \ddots & \ddots & & & & & & \\
     & & & & & & & a_{m-1}^{p-1} & b_{m-1}^{p-1} & c_{m-1}^{p-1} & & & & & \\
     & & & & & & & & a_m^{p-1} & b_m^{p-1} & \boxed{c_m^{p-1}} & & & & \\
     & & & & & & & & & \boxed{a_1^p} & b_1^p & c_1^p & & & \\
     & & & & & & & & & & a_2^p & b_2^p & c_2^p & & \\
     & & & & & & & & & & & \ddots & \ddots & \ddots & \\
     & & & & & & & & & & & & a_{m-1}^p & b_{m-1}^p & c_{m-1}^p \\
     & & & & & & & & & & & & & a_m^p & b_m^p 
  \end{pmatrix}.
\end{equation}
Let $e_i$ be a column vector with its $i\mathrm{th}~(1 \leq i \leq n)$ element being one and all the other entries being 
zero. The matrix $\Delta A$ can be written as
\begin{equation}\label{eq13}
  \Delta A = 
  \begin{bmatrix}
    a_{m+1}e_{m+1}, c_me_m, \dots, a_{(p-1)*m+1}e_{(p-1)*m+1}, c_{(p-1)*m}e_{(p-1)*m}
  \end{bmatrix}
  \begin{bmatrix}
    e_m^T \\
    e_{m+1}^T \\
    \vdots \\
    e_{(p-1)*m}^T \\
    e_{(p-1)*m+1}^T
  \end{bmatrix}
  = VE^T,
\end{equation}
where both $V$ and $E$ are $n \times 2(p-1)$ matrices. Hence, the matrix $A$ can be factored into
\begin{equation}\label{eq14}
  A = \tilde{A} + VE^T.
\end{equation}
In this way, equation (\ref{eq10}) can be solved using the Sherman-Morrison-Woodbury formula\cite{Sherman1950,Woodbury1950}, that is
\begin{equation}\label{eq15}
  x = A^{-1}d = (\tilde{A}+VE^T)^{-1}d = \tilde{A}^{-1}d - \tilde{A}^{-1}V (I + E^T\tilde{A}^{-1}V)^{-1} E^T \tilde{A}^{-1} d.
\end{equation}
Let $Z=I+E^T Y$, then equation (\ref{eq15}) can be solved step by step as follows
\begin{align}
  \tilde{A}\tilde{x} &= d, \label{eq16} \\
  \tilde{A}Y &= V, \label{eq17} \\
  Zy &= E^T \tilde{x}, \label{eq18} \\
  x &= \tilde{x} - Yy. \label{eq19}
\end{align}

Since the coefficient matrix is the same, both equation (\ref{eq16}) and equation (\ref{eq17}) can be solved using the 
Thomas algorithm
\begin{equation}\label{eq20}
  A_i \left[ \tilde{x}^i, v^i, w^i \right] = \left[ d^i, a_1^i e_1, c_m^i e_m \right], (1 \leq i \leq p, a_1^1 = c_m^p = 0),
\end{equation}
where $v^i$ and $w^i$ are potentially nonzero column vectors of the ith row block of the matrix $Y$. Here the length of 
column vector $e$ is $m$. The solution can be performed completely in parallel.

The solution of equation (\ref{eq18}) is the core of the PDD algorithm. The matrix $Z$ has the following form
\begin{equation}\label{eq21}
  Z = 
  \begin{pmatrix}
    1 & w_m^1 & 0 & & & & & \\
    v_1^2 & 1 & 0 & \boxed{w_1^2} & & & & \\
    \boxed{v_m^2} & 0 & 1 & w_m^2 & 0 & & & \\
    & \ddots & \ddots & \ddots & \ddots & \ddots & & \\
    & & \ddots & \ddots & \ddots & \ddots & \ddots & \\
    & & & 0 & v_1^{p-1} & 1 & 0 & \boxed{w_1^{p-1}} \\
    & & & & \boxed{v_m^{p-1}} & 0 & 1 & w_m^{p-1} \\
    & & & & & 0 & v_1^p & 1 \\
  \end{pmatrix}.
\end{equation}
Sun et al.\cite{Sun1989} points out that the magnitude of the last component of $v^i$, i.e. $v_m^i$, and the first component of 
$w^i$, i.e. $w_1^i$, may be smaller than machine accuracy when $p \ll n$ (e.g., $ n/p \geq 50$), especially for diagonal 
dominant tridiagonal systems. In this case, $v_m^i$ and $w_1^i$ can be dropped (the entries with frame shown in equation 
(\ref{eq21})), and $Z$ becomes a block diagonal system consisting of $(p-1)$ independent $2\times2$ blocks. Hence, 
equation (\ref{eq18}) can be solved efficiently in parallel, which leads to the highly efficiently PDD algorithm.

The PDD algorithm, using $p$ processors, consists of the following steps:
\begin{enumerate}
  \item Allocate $A_i, d^i, a_1^i$ and $c_m^i$ to the $i$th processor, where $1 \leq i \leq p$.
  \item Solve equation (\ref{eq20}) in parallel on p processors, using the Thomas algorithm.
  \item Send $\tilde{x}_1^i, v_1^i$ from the $i$th processor to the $(i-1)$th processor, where $2 \leq i \leq p$.
  \item On the first $(p-1)$ processors, solve the following equation in parallel using Cramer's Rule
        \begin{equation}
          \begin{pmatrix}
            1 & w_m^i \\ 
            v_1^{i+1} & 1
          \end{pmatrix}
          \begin{bmatrix}
            y_1^i \\ y_2^i
          \end{bmatrix}
          =
          \begin{bmatrix}
            \tilde{x}_m^i \\ \tilde{x}_1^{i+1}
          \end{bmatrix},
          (1 \leq i \leq p-1),
        \end{equation}
        where $y_j^i = y_{(i-1)\times2+j}, j=1,2$. Then send $y_1^i$ from the $i$th processor to the $(i+1)$th processor.
  \item Calculate equation (\ref{eq19}) in parallel on all processors, and the final solution is obtained.
        \begin{equation}
          x^i = \tilde{x}^i - \left[ v^i, w^i \right] \begin{bmatrix} y_1^{i-1} \\ y_2^i \end{bmatrix}, (1 \leq i \leq p, y_1^0 = y_2^p = 0).
        \end{equation}
\end{enumerate}

\begin{figure}[ht]
  \centering
  \includegraphics[width=175pt]{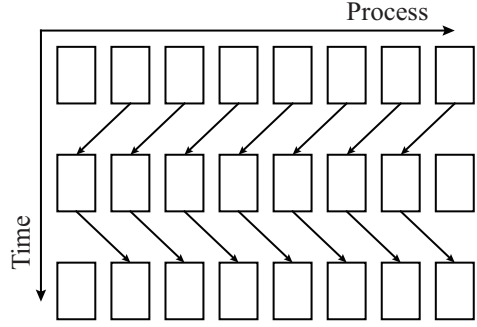}
  \caption{The communication pattern of the PDD algorithm\cite{Sun1989}.\label{fig3}}
\end{figure}

It can be seen that only two sendrecv communications are required when using the PDD algorithm to solve a tridiagonal 
linear system, which contributes to the high parallel efficiency. The communication pattern is illustrated in Figure 
\ref{fig3}. It is pointed out in Ref.\cite{Sun1995} that for diagonally dominant tridiagonal systems, the PDD algorithm 
approximates the true solution to within machine accuracy when the size of parallel partitions satisfies certain requirements.

\section{Parallel strategy and implementation}\label{sec3}

Using an explicit time marching scheme, the update of velocity fields in the projection method is easy to parallelize. 
However, the global coupling of the pressure Poisson equation is often a critical challenge to massively parallel 
numerical simulations. Hence, this section focuses on the parallel strategy and the efficient implementation for the 
pressure Poisson equation.

\subsection{Domain decomposition}

Domain decomposition is one of the most important implementations of data parallelism in CFD. The computational domain 
for wall turbulence is usually a rectangular box. Three decomposition schemes are available, i.e. 1D plate-like 
decomposition, 2D pencil-like decomposition and 3D cube-like decomposition. The adoption of the final decomposition 
scheme should be considered carefully.

On the one hand, the parallel scalability and the communication overhead are affected by the decomposition. Consider 
the parallel decomposition of a cubic computational domain with periodic boundary conditions. The amount of computational 
work of each subdomain is proportional to the volume of itself, while the amount of communication required is proportional 
to the subdomain surface area. Then the computation-to-communication ratio, reflecting the parallel efficiency to some 
extent, roughly equals to the subdomain volume-to-area ratio. It should be maximized for good scalability and high 
parallel efficiency. It is not hard to find that both 2D pencil-like decomposition and 3D cube-like decomposition have 
a higher volume-to-area ratio, thus leading to less communication overhead and higher parallel efficiency. 

On the other hand, the choice of decomposition is often limited by the adopted numerical method. The FFT calculation 
involved in solving the pressure Poisson equation is a global operation, which requires that each subdomain not be 
decomposed in the direction of transformation execution. If a 3D cube-like decomposition is adopted, then it should 
be reduced to a 2D pencil-like decomposition before performing FFTs, which results in extra overhead of data movement 
and communication.

From the above, a 2D pencil-like decomposition is chosen for the parallelization implementation of the numerical 
method, considering the trade-off between enhancing scalability and accommodating global algorithm features.

\subsection{A parallel method for solving the pressure Poisson equation}

The 2D FFT calculation in the horizontal directions is the first step to solve the pressure Poisson equation. Using a 
2D pencil-like decomposition, FFTs in the long direction of the pencil-like subdomain can be performed in parallel. 
Hence, an x-pencil decomposition should be the initial decomposition to facilitate FFTs in $x$ direction. Then an 
all-to-all global transpose from x-pencil decomposition to y-pencil decomposition is required to perform FFTs in $y$ 
direction.

\begin{figure}[htb]
  \centerline{\includegraphics[height=110pt]{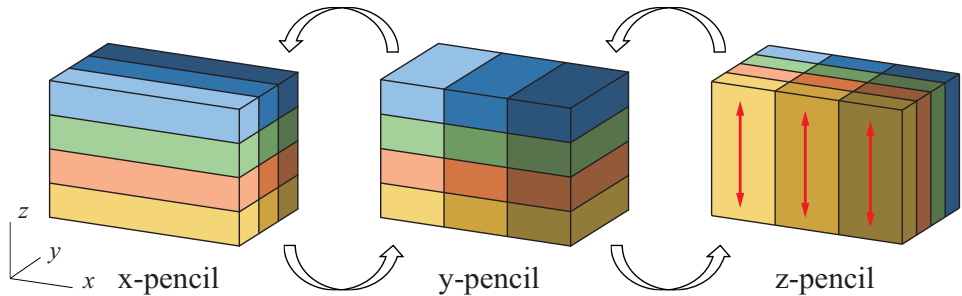}}
  \caption{
    Illustration of 2D pencil-like domain decomposition and transpose in the 
    process of solving the pressure Poisson equation using Thomas algorithm. 
    The computational domain is decomposed into $3\times4$ subdomains, which 
    is distinguished by a different color. The curved arrow indicates the 
    transpose. The double-headed straight arrow indicates the solving process 
    of Thomas algorithm.  
  \label{fig4}}
\end{figure}

\begin{figure}[htb]
  \centerline{\includegraphics[height=110pt]{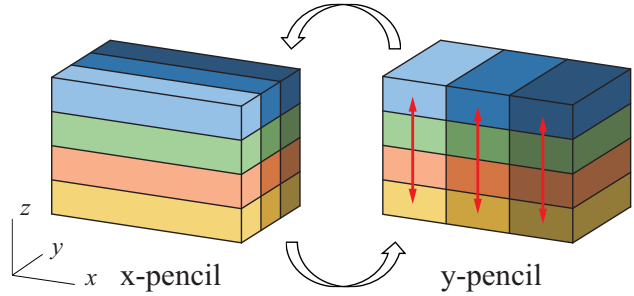}}
  \caption{
    Illustration of 2D pencil-like domain decomposition and transpose in the 
    process of solving the pressure Poisson equation using PDD algorithm. The 
    double-headed straight arrow indicates the solving process of PDD algorithm.  
  \label{fig5}}
\end{figure}

After a forward 2D FFT, a series of tridiagonal linear equations merely coupled in $z$ direction are obtained. However, 
due to the y-pencil decomposition at the moment, the equations cannot be solved directly using the Thomas algorithm. 
A common solution is to transpose y-pencil decomposition to z-pencil decomposition, ensuring all required data in $z$ 
direction contained in each subdomain, as illustrated in Figure \ref{fig4}. In this case, the whole solution requires 
four all-to-all global transposes. It is worth noting that the communication overhead of the all-to-all transpose is 
often expensive, in practice more time-consuming than the FFT operation. The transpose process will likely become a 
bottleneck of parallel performance in a massively parallel simulation. Hence, the PDD algorithm with low communication 
overhead is adopted to solve the tridiagonal systems in parallel, reducing two transposes between y-pencil decomposition 
and z-pencil decomposition. The solving process is illustrated in Figure \ref{fig5}.

In summary, the parallel method for solving the pressure Poisson equation is as follows:

\renewcommand{\labelenumi}{(\theenumi)}
\begin{enumerate}
  \item Compute the right hand side (RHS) of the pressure Poisson equation, in x-pencil decomposition.
  \item Perform forward 1D FFTs in $x$ direction on the result of (1), in x-pencil decomposition.
  \item Transpose the result of (2) from x-pencil decomposition to y-pencil decomposition.
  \item Perform forward 1D FFTs in $y$ direction on the result of (3), in y-pencil decomposition.
  \item Solve the tridiagonal systems in parallel using the PDD algorithm, in y-pencil decomposition.
  \item Perform forward 1D FFTs in $y$ direction on the result of (5), in y-pencil decomposition.
  \item Transpose the result of (6) from y-pencil decomposition to x-pencil decomposition.
  \item Perform forward 1D FFTs in $x$ direction on the result of (7), in x-pencil decomposition.
\end{enumerate}

The method reduces the number of all-to-all transposes to two, and only introduces two sendrecv communications between 
neighboring subdomains, whose overhead is much smaller than that of all-to-all. Although the PDD algorithm requires 
some additional computation, the cost can be offset by the performance gains in communication.

\subsection{Efficient implementation}

The final parallel performance of the code depends not only on the parallel algorithm, but also on the implementation 
method. The code is developed in FORTRAN90/95. Considering that mainstream large-scale computing clusters all adopt a 
hybrid architecture of distributed memory and shared memory, the MPI/OpenMP hybrid programming model is utilized to 
give full play to the computing power of the machine and improve the parallel performance of the solver. The MPI 
process, corresponding with the decomposition subdomain, is responsible for calculation and communication at a coarse-
grained parallel level. The OpenMP thread mainly implement fine-grained parallelization of the calculation part.

Communications between adjacent subdomains are necessary when calculating the spatial discretization for the convective, 
viscous, pressure terms, etc. For the second order centered finite difference adopted in this paper, the number of 
stencil points is three. Hence, an additional layer of halo cells should be set up around the subdomain to store a 
copy of boundary cells data of adjacent subdomains. Using 2D pencil-like decomposition, four pairwise data exchanges 
is required per halo update, which can be realized by calling the MPI\_SENDRECV function.

\begin{figure}[htb]
  \centerline{\includegraphics[width=350pt]{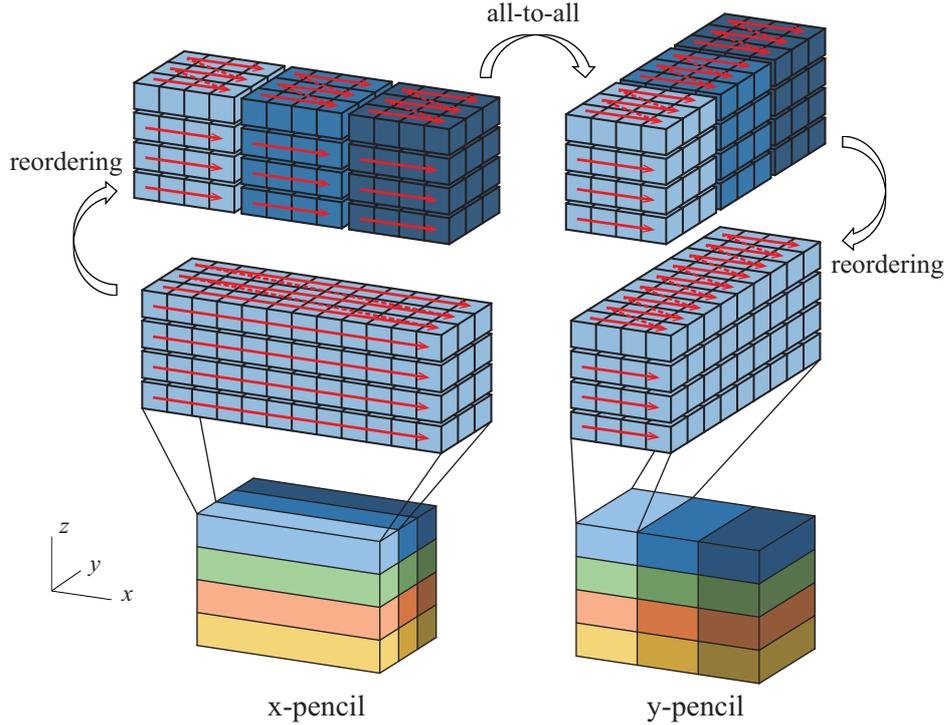}}
  \caption{
    Illustration of data reordering in the transpose from x-pencil decomposition 
    to y-pencil decomposition. The straight arrow and dotted line indicate the 
    contiguous memory access order. Data sent to or received from different 
    processes are distinguished by different colors.
  \label{fig6}}
\end{figure}

Both FFT operation and data global transpose are critical steps of the parallel method for solving the pressure 
Poisson equation. In order to obtain good parallel performance, a multi-dimensional FFT and transpose module which 
supports MPI/OpenMP hybrid parallelism is developed by referring to the implementation of several open source libraries 
(e.g. PFFT\cite{PFFT}, P3DFFT\cite{P3DFFT}, 2DECOMP\&FFT\cite{LiLaizet2010}, etc.), which is highly efficient and scalable. Firstly, it is straight-forward to 
perform FFTs on batch data in parallel by using multiple threads, since the data lines to be transformed are independent 
of each other. It can be easily implemented by calling the multithreaded computing interface of general FFT libraries 
(e.g. FFTW\cite{FFTW3}, MKL\cite{IntelMKL}, etc.). Secondly, data reordering before and after the all-to-all transpose can also benefit 
from multithreaded parallelism. Take the transpose from x-pencil decomposition to y-pencil decomposition as an example, 
data reordering as shown in Figure \ref{fig6} is required, before and after the all-to-all transpose. In other words, data sent 
to (or received from) different processes should be aggregated into a continuous memory buffer. The speed of this 
reordering is clearly limited by the memory bandwidth, as nothing but memory access is occurring. Note that data 
reordering is independent in $z$ direction and can be processed in parallel using multiple threads, which can improve the 
memory bandwidth utilization by maintaining multiple data streams. Finally, the performance of the all-to-all 
communication mainly depends on the underlying implementation of the MPI library and the network architecture of the 
computing system. By adjusting communication parameters of the MPI library, the communication bandwidth can also be 
improved to some extent.

As for the parallel implementation of the PDD algorithm, the method of ``batch calculation, batch communication'' is 
adopted to reduce the communication overhead. After completing the forward 2D FFT calculation, adjacent processes in 
$z$ direction need to jointly solve multiple (assuming $n$) mutually independent tridiagonal linear equations. If solutions 
are performed sequentially, on the one hand, $2n$ sendrecv communications are required according to the introduction of 
the PDD algorithm in Section \ref{sec2}. Since small messages are transmitted, the communication overhead is dominated by the 
communication delay. On the other hand, it may also lead to a large number of strided memory accesses and reduce the 
computing performance. For this reason, the solution proposed in this paper is: each process uses multithreading to 
complete the pure computing tasks of solving tridiagonal systems at one time, aware of the locality principle. The 
calculation results are aggregated into a continuous memory buffer by using the derived datatype in MPI, and then 
transmitted in the form of large messages. This method can reduce the number of sendrecv communications to two, which 
minimizes the communication overhead and improves the communication bandwidth.

\subsection{PowerLLEL}

From the above, several primary parallel strategies and optimizations are adopted to efficiently implement the 
proposed numerical method. The 2D pencil-like domain decomposition ensures the good scalability of the code, while 
the flexible MPI/OpenMP hybrid programming model speeds up not only the calculation of the momentum equation but 
also the PDD-based solution to the pressure Poisson equation. Finally, a new parallel solver for the 3D incompressible 
Navier-Stokes equations, PowerLLEL, is developed. It has the significant potential for massively parallel DNS of wall 
turbulence at a high Reynolds number and can be easily applied to the simulation of other incompressible flows.

\section{Parallel performance}\label{sec4}

In this section, the parallel performance of PowerLLEL is presented through a series of strong and weak scalability 
tests. The performance difference between two parallel running modes, namely pure MPI and MPI/OpenMP hybrid, is 
also discussed. All tests are performed on the CPU partition of the Tianhe-2A supercomputer, currently ranked 
No. 6 in the world\cite{Top500}. Each computing node contains dual 12-core Intel Xeon E5-2692 v2 processors and 64 GB of 
memory. A dedicated network TH Express-2 based on the fat tree topology is adopted for node interconnection. The 
software environment for test is shown in Table \ref{tab1}. The grid configuration is shown in Table \ref{tab2}. 
Both in the strong and weak scalability tests (denoted by S and W respectively), two sets of grids of different 
sizes are used.

\begin{table}[ht]
  \centering
  \caption{Software environment for scalability tests.\label{tab1}}
  \small
  \begin{tabular}{ll}
  \toprule
  \textbf{Item} & \textbf{Name (version)} \\
  \midrule
  OS & Red Hat (6.5)  \\
  Compiler & Intel Fortran Compiler (14.0.2)  \\
  MPI Library & MPICH (3.1.3)  \\
  Math Library & FFTW (3.3.8)  \\
  IO Library & HDF5 (1.10.4)  \\
  \bottomrule
  \end{tabular}
\end{table}

\begin{table}[ht]
  \centering
  \caption{Grid configurations for scalability tests. Characters ``S'' and ``W'' in 
  the item ``Case'' indicate the strong scalability test and weak scalability test, 
  respectively. $N_x$, $N_y$, $N_z$ are the number of points in $x, y, z$ directions.
  \label{tab2}}
  \small
  \begin{tabular}{lllll}
  \toprule
  \textbf{Case} & $N_x$ & $N_y$ & $N_z$ & \textbf{Grid points} \\
  \midrule
  S1 & 2304 & 1536 & 864 & $3.06 \times 10^9$ (Total)  \\
  S2 & 4608 & 3072 & 1152 & $16.3 \times 10^9$ (Total)  \\
  W1 & 1152/2304/4608 & 768/1536/3072 & 864 & $2.65 \times 10^6$ (Per core)  \\
  W2 & 1536/3072/6144/9216 & 1152/2304/4608 & 864 & $5.31 \times 10^6$ (Per core)  \\
  \bottomrule
  \end{tabular}
\end{table}

\begin{figure}[ht]
  \centering
  \includegraphics[]{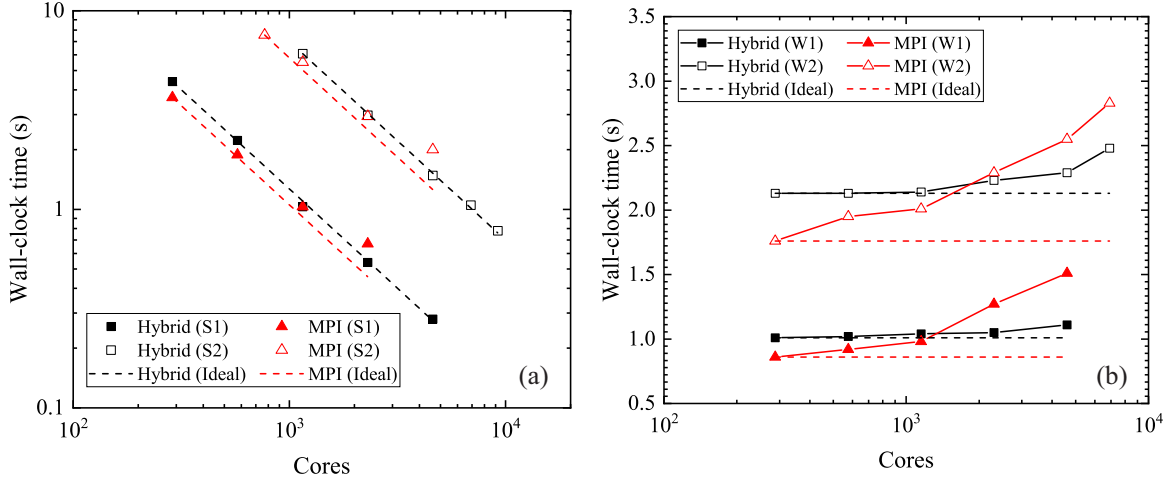}
  \caption{
    Strong (a) and weak (b) scalability tests for PowerLLEL. The dashed lines indicate 
    the ideal scaling behavior. Performances in two parallel running modes, 
    namely pure MPI and MPI/OpenMP hybrid, are labeled with red triangles and 
    black squares respectively.
  \label{fig7}}
\end{figure}

As illustrated in Figure \ref{fig7}(a), PowerLLEL in MPI/OpenMP hybrid mode shows excellent strong scalability up to 
9216 cores for a grid with 16.3 billion points, while it performs poor in pure MPI mode. The main reason is that as 
the number of CPU cores increases, the communication overhead in pure MPI mode increases faster, which leads to the 
decrease of parallel efficiency. In the case of weak scaling, the overall grid size becomes larger but the workload 
of each core remains the same. Linear scaling is achieved if the run time stays constant. Actually, the pure calculation 
time should be almost the same and only the increasing communication overhead will lead to the reduction of parallel 
efficiency. As shown in Figure \ref{fig7}(b), with the increase of cores, only a slight increase of the wall-clock 
time is observed in MPI/OpenMP hybrid mode, which suggests that PowerLLEL is able to solve a larger problem (i.e. 
higher Reynolds number) on a larger-scale system with a slight performance penalty. Moreover, the rapid increase of 
wall-clock time in pure MPI mode once again demonstrates that the communication overhead is the main reason for the 
relatively low parallel efficiency.

Finally, the absolute performance of PowerLLEL is also excellent. No matter in which mode, the wall-clock time per 
timestep of the code is very small, especially in the large-scale parallel numerical simulations. It is noteworthy 
that when the number of cores is small and the communication overhead has not become a bottleneck, the overall code 
performance in the pure MPI mode is better than that in the MPI/OpenMP hybrid mode.

\section{Validation}\label{sec5}

\begin{table}[hb]
  \centering
  \caption{Summary of simulation parameters for turbulent channel flow cases. 
  The initial letter of each case indicates the size of the domain: small (S), 
  medium (M) and large (L). $L_x$ and $L_z$ are the computational domain sizes 
  in the streamwise and spanwise directions. $N_x, N_y, N_z$ are the number of 
  grid points in the streamwise, wall-normal and spanwise directions. 
  $\Delta x^+, \Delta y^+, \Delta z^+$ are the corresponding grid spacings. The 
  statistical averaging time, $T_{\mathrm{stat}}$, is given in terms of the eddy 
  turn over time $\delta/u_{\tau}$.
  \label{tab3}}
  \small
  \begin{tabular}{lllllllll}
  \toprule
  \textbf{Case} & $Re_{\tau}$ & $L_x/\delta \times L_z/\delta$ & $N_x \times N_y \times N_z $ & $\Delta x^+$ & $\Delta z^+$ & $\Delta y^+$ & $T_\mathrm{stat} u_{\tau}/\delta$ & \textbf{Line style} \\
  \midrule
  L550  &  542 & $4\pi \times 2\pi$ & 1024$\times$360$\times$512  & 6.65 & 6.65 & 0.20--7.63 &  8.13 & \textcolor{black}{------} (black)  \\
  L1000 &  991 & $4\pi \times 2\pi$ & 2048$\times$540$\times$1024 & 6.08 & 6.08 & 0.21--9.70 &  3.96 & \textcolor{blue}{------} (blue) \\
  L2000 & 2003 & $4\pi \times 2\pi$ & 4096$\times$864$\times$2048 & 6.14 & 6.14 & 0.23--12.6 &  3.41 & \textcolor{red}{------} (red) \\
  M550  &  542 & $2\pi \times  \pi$ &  512$\times$360$\times$512  & 6.65 & 3.32 & 0.15--6.35 & 10.75 & \textcolor{black}{-- -- --} (black)  \\
  S1000 &  991 & $ \pi \times\pi/2$ &  384$\times$600$\times$384  & 8.11 & 4.05 & 0.17--6.96 & 11.89 & \textcolor{blue}{-- -- --} (blue) \\
  \bottomrule
  \end{tabular}
\end{table}

To validate the simulation results of PowerLLEL, five new DNSs of the channel flow are carried out at friction Reynolds 
numbers $Re_{\tau} = u_{\tau}\delta / \nu$ = 550, 1000, 2000, where $u_{\tau} = \sqrt{\tau_w/\rho}$ is the 
friction velocity, $\tau_w$ is the wall shear stress, $\delta$ is the channel half-height and $\nu$ is the kinematic 
viscosity coefficient. For the sake of convenience, $x$, $y$ and $z$ here indicate the streamwise, wall-normal and 
spanwise directions, respectively. Three simulations are computed in a large domain with streamwise and spanwise sizes 
of $L_x=4\pi\delta$ and $L_z=2\pi\delta$, while the other two use smaller boxes with $L_x=2\pi\delta$, $L_z=\pi\delta$ 
and $L_x=\pi\delta$, $L_z=0.5\pi\delta$, respectively. A minimal channel has been proved to provide correct one-point 
statistics and can be employed as an economical alternative to the costly large-box channel flow DNS\cite{LJ2014,Flores2010,Jimenez1991}; this 
has also been verified here, indicating the reliable performance of PowerLLEL. Hence, the two small-box simulations 
can lay a foundation for DNS of channel flow at a higher Reynolds number. Uniform grids are used in the wall-parallel 
directions while a non-uniform grid clustered towards the wall is instead used in the wall-normal direction. Details 
on the grid configuration are provided in Table \ref{tab3}. The simulation of case L550 is initiated with a parabolic 
streamwise velocity profile with superposed random perturbations. The initial field of case L1000 at a higher $Re_\tau$ 
is interpolated from the fully developed field of case L550. The rest can be done in a similar fashion. Statistics are 
collected once the friction velocity $u_\tau$ reaches a statistically steady state. The reference data comes from the 
very recent simulations by Lee \& Moser\cite{LM2015}, which has been shown to be with high quality via various means (see, for 
example, in Ref.\cite{Chen2019}). The lines in Table \ref{tab3} are used consistently in later plots, unless otherwise stated.

\begin{figure}
  \centering
  \includegraphics[]{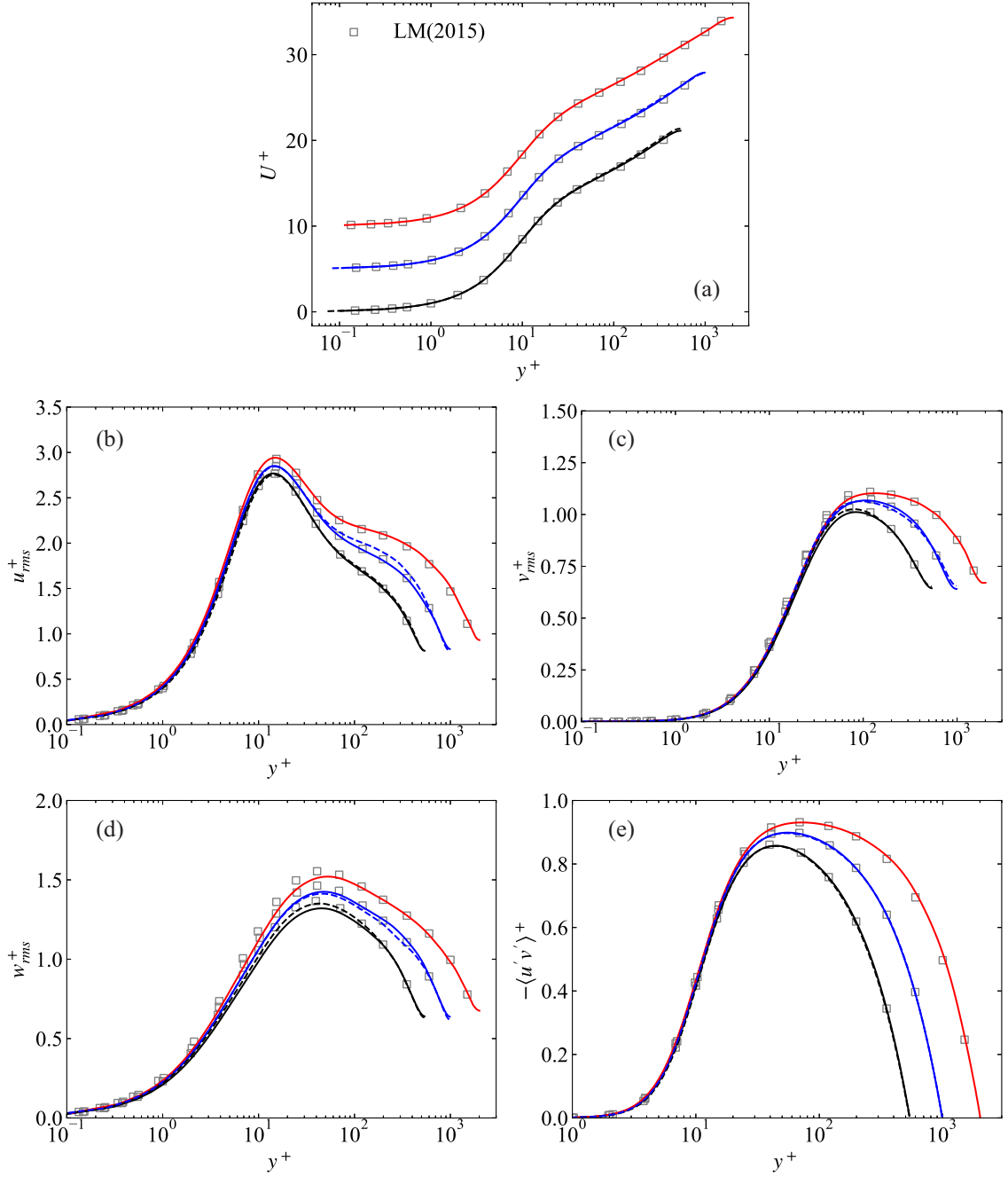}
  \caption{
    Profiles of the mean streamwise velocity (a), the turbulent velocity 
    fluctuations (b, c, d) and the Reynolds shear stresses (e). Profiles of 
    the mean streamwise velocity at $Re_{\tau}$ = 1000 and $Re_{\tau}$ = 2000 
    are offset by 5 and 10 units, respectively. Data from Lee et al.\cite{LM2015} 
    are marked with hollow squares, consistent in later plots unless otherwise 
    stated. Lines are our simulations as specified in Table \ref{tab3}.
  \label{fig8}}
\end{figure}

Profiles of the mean streamwise velocity are shown in Figure \ref{fig8}(a). It can be seen that the DNS results in this paper 
are in good agreement with the reference data and present obviously a common logarithmic behavior for all the Reynolds 
numbers and box sizes. As presented in Figure \ref{fig8}(b)--\ref{fig8}(d), the turbulent velocity fluctuation 
profiles also agree reasonably with the reference data, further illustrating that it is feasible to introduce the PDD-based 
approximate solver for the pressure Poisson equation to DNS of the channel flow.

\begin{figure}
  \centering
  \includegraphics[]{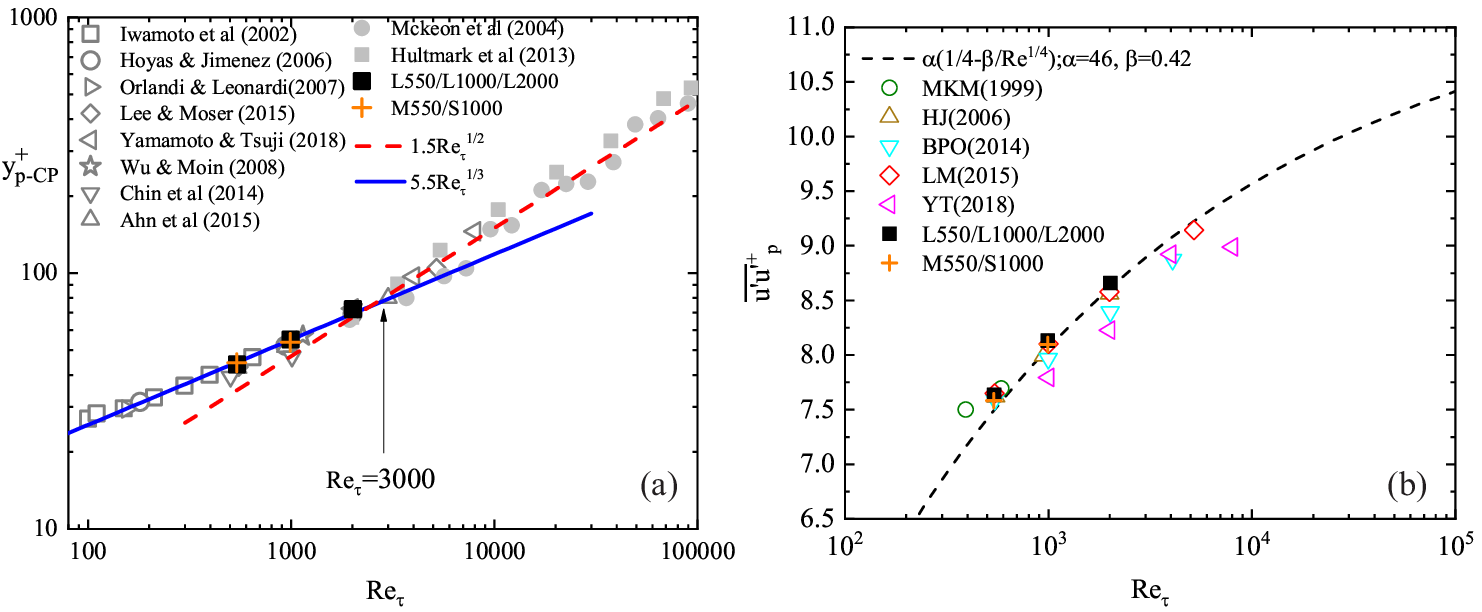}
  \caption{
    (a) Re-scaling of the peak location of $-\langle u'v' \rangle^+$;
    (b) Re-scaling of the streamwise turbulence intensity peak $\langle u'u' \rangle_p^+$.
    In (a), the solid line indicates the $Re_{\tau}^{1/3}$ scaling for 
    small $Re$ proposed by Chen et al.\cite{Chen2019}; the dashed line indicates 
    the $Re_{\tau}^{1/2}$ scaling for high $Re$; the grey symbols are 
    previous DNS/experimental data compiled in Ref.\cite{Chen2019}; the solid 
    squares indicate the present DNSs in a large box; the plus signs 
    indicate the present DNSs in a smaller box.
    In (b), the dashed line indicates the $Re$ scaling proposed by Chen et al.\cite{Chen2020};
    the hollow symbols are previous DNS data of the channel flow\cite{MKM1999,HJ2006,BPO2014,LM2015,Yamomoto2018}.
  \label{fig9}}
\end{figure}

The profiles of Reynolds shear stress at different Reynolds numbers are illustrated in Figure \ref{fig8}(e), in good agreement 
with the reference data. Note that it is believed for a long time that the peak location $y_p^+$ of $-\langle u'v' \rangle^+$ 
follows a $Re_{\tau}^{1/2}$ power law due to the logarithmic mean velocity profile. However, recent studies by 
Chen et al.\cite{Chen2019} show that there is a scaling transition for the peak location (Figure \ref{fig9}(a)). That is, for 
small $Re$, the peak location is in the buffer layer rather than in the logarithmic layer, which thus nullifies the previous 
derivation of the $Re_{\tau}^{1/2}$ scaling (valid for large $Re_{\tau}$). Instead, Chen et al.\cite{Chen2019} presents an 
alternative $Re_{\tau}^{1/3}$ scaling for small $Re_{\tau}$ (less than 3000) based on the buffer layer eddies 
represented by the characteristic length function. In the Figure \ref{fig9}(a), we show our simulations together with previous 
DNS data, all in close agreement with the new $Re_{\tau}^{1/3}$ scaling, demonstrating well the quality of our simulations.

As for the streamwise turbulence intentsity peak $\langle u'u' \rangle_p^+$, the existing explanations suggest a logarithmic 
growth with respect to $Re_{\tau}$. Recently, Chen \& Sreenivasan\cite{Chen2020} propose an alternative formula for the peak
magnitude that approaches a finite limit owing to the natural constraint of boundedness on the dissipation rate at the wall.
As shown in Figure \ref{fig9}(b), our simulations together with previous DNS data of the channel flow validate this new formula.

\begin{figure}
  \centering
  \includegraphics[]{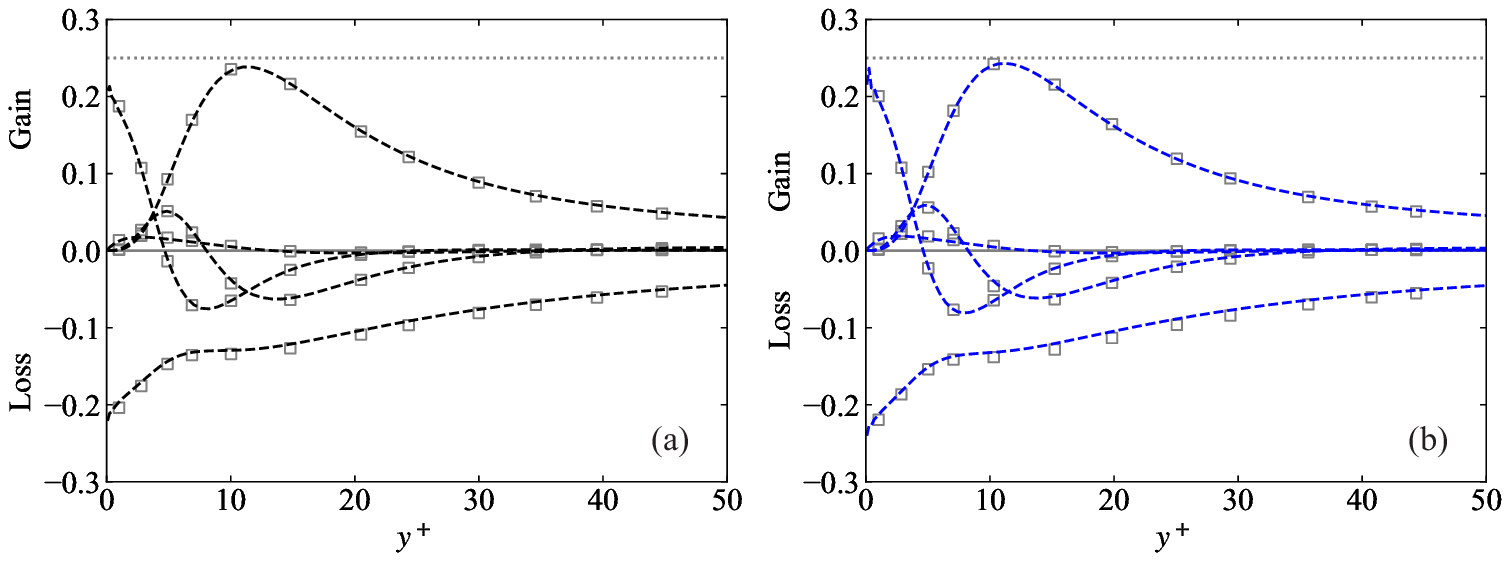}
  \caption{
    Budget of turbulent kinetic energy. (a) $Re_{\tau}$ = 550; (b) $Re_{\tau}$ = 1000. 
    The position of 1/4 is marked with a black dot-dashed line for reference.
  \label{fig10}}
\end{figure}

In Figure \ref{fig10}, the budget of turbulent kinetic energy is plotted, comparing well with the reference data. All the cases 
show that the dominant balance is between production and dissipation away from the wall for $y^+ \gtrapprox 10$, while 
near the wall, it is the diffusion balancing the dissipation. Pressure transport is effective only in the domain 
$y^+ \lessapprox 10$, and turbulent convection shows a positive energy gain for $y^+ \lessapprox 7$ while redistributes 
energy outwards for larger $y^+$. Note that all the productions show the maximum value 1/4 at $y^+ \approxeq 12$, 
which has been explained, e.g. in Cantwell\cite{Cantwell2019}. Moreover, towards the centerline, production becomes zero due to the 
vanishing mean shear and it is the convection balancing dissipation near the center\cite{Chen2018}; this is also observed in our 
simulations but not shown here.

\begin{figure}
  \centering
  \includegraphics[]{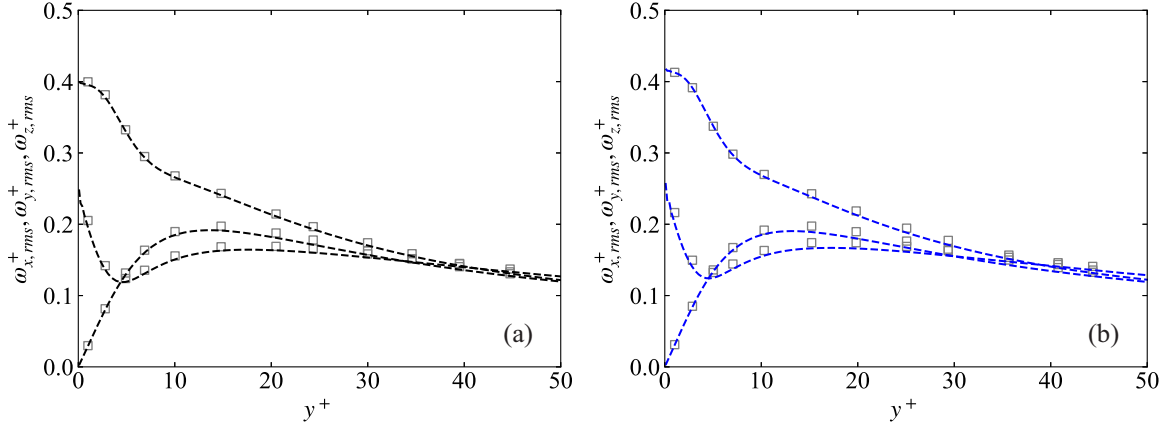}
  \caption{
    Profiles of root-mean-square (rms) of vorticity fluctuations. (a) $Re_{\tau}$ = 550; (b) $Re_{\tau}$ = 1000.
  \label{fig11}}
\end{figure}

The wall-normal dependence of vorticity fluctuations is shown in Figure \ref{fig11}. At the wall, $\omega_{y,rms}^+$ is zero due to the 
non-slip wall condition; $\omega_{z,rms}^+$ is the largest caused by the shear $\mathrm{d}u/\mathrm{d}y$, and $\omega_{x,rms}^+$ 
is also nonzero due to the effect of streamwise vortices. Away from the wall, $\omega_{y,rms}^+$ increases rapidly due 
to the streaks ($\mathrm{d}u/\mathrm{d}z \neq 0$), and for $y^+$ larger than 40, all the root-mean-square (rms) of 
vorticities are about the same value, indicating the turbulent flow is more isotropic than near the wall. In the above 
plots, our simulations are again in close agreement with the reference data.

\begin{figure}
  \centering
  \includegraphics[]{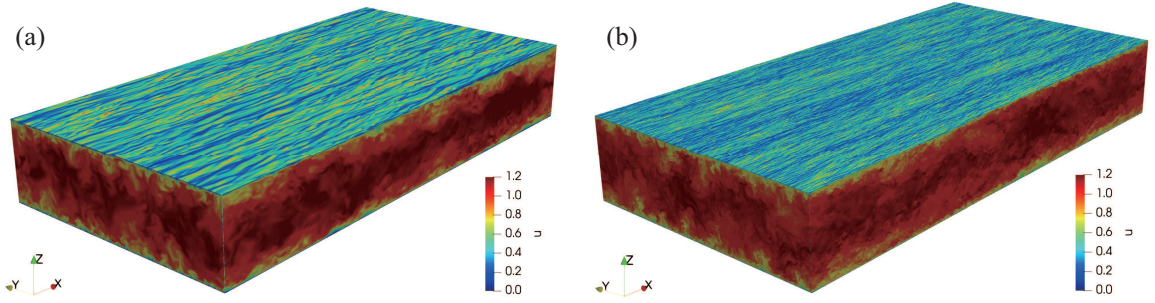}
  \caption{
    Instantaneous streamwise velocity for turbulent channel flow at $Re_{\tau}$ = 550 (a) 
    and $Re_{\tau}$ = 2000 (b). The wall-parallel plane is at $y^+$ = 10.
  \label{fig12}}
\end{figure}

Finally, Figure \ref{fig12} shows a contour plot of the instantaneous streamwise velocity in a cross-stream, a streamwise and a 
wall-parallel plane at $y^+$ = 10. Low- and high-speed streaks, typical flow organization of incompressible wall turbulence, 
are distinctly shown in the wall-parallel plane. Sweep and ejection events can also be observed in the cross-stream 
plane. Compared to the case at $Re_{\tau}$ = 550, there are more streaks distributed along the spanwise direction at 
$Re_{\tau}$ = 2000; however, their average spacing $z^+ \approxeq 100$ is almost invariant, consistent with previous 
observations\cite{Marusic2010,Jimenez2012}. Reducing the intensity of these streak structures would lead to a significant reduction of 
friction drag at the wall, as pursued by many recent studies\cite{Yao2018}.

\section{Conclusions and prospects}\label{sec6}

In this paper, a highly efficient parallel method for solving the 3D incompressible Navier-Stokes equations is developed. 
The spatial discretization scheme is the standard second-order centered finite difference while an explicit second-order 
Runge-Kutta scheme is used for the time advancement. The core of the current method is an approximate solver for the 
pressure Poisson equation, which combines 2D FFTs with a highly efficient parallel approximate algorithm for tridiagonal 
systems, namely the PDD algorithm. Using a 2D pencil-like domain decomposition, only two all-to-all global communications 
and two sendrecv adjacent communications are involved in the process of solving the pressure Poisson equation in parallel. 
The proposed method is efficiently implemented in the MPI/OpenMP hybrid programming model and the 2D pencil-like domain 
decomposition, and a high-performance solver (i.e. PowerLLEL) is eventually developed. A series of the strong scalability 
tests have been carried out on the Tianhe-2A supercomputer. The results indicate that PowerLLEL in the MPI/OpenMP hybrid 
mode shows excellent strong scalability up to $10^4$ cores when the number of grid points is 16.3 billion, and the wall-clock 
time per timestep is very small. Several direct numerical simulations of the channel flow at different friction Reynolds 
numbers ($Re_{\tau}$ = 550, 1000, 2000) have been performed to validate the reliability of the proposed method, showing good 
agreement with reference data. With the help of the proposed method and PowerLLEL, it is possible to carry out massively 
parallel DNS of wall turbulence at higher Reynolds number with relatively low computational cost will be performed in future.

Note that PowerLLEL can also be applied for DNS of other incompressible flows (e.g. turbulent boundary layer, pipeline flow, 
etc.). Combined with the immerse boundary method, it is capable to simulate flows with complex boundary conditions (such as 
the simulation of boundary layer with surface roughness), to be pursued in future. Finally, it is our next goal to migrate 
PowerLLEL from pure CPU systems to the CPU-GPU heterogeneous systems.

\addcontentsline{toc}{section}{Acknowledgements}
\section*{Acknowledgments}
This work was supported by the National Natural Science Foundation of China (Grant No. 11772362). The authors would like to 
thank the computational sources provided by the National Supercomputer Center in Guangzhou.

\addcontentsline{toc}{section}{Disclosure statement}
\section*{Disclosure statement}
No potential conflict of interest was reported by the author(s).

\addcontentsline{toc}{section}{Reference}
\bibliography{ref}

\begin{thebibliography}{10}
\providecommand \doibase [0]{http://dx.doi.org/}%

\bibitem{Abide2017}
Abide S, Binous MS, Zeghmati B. An efficient parallel high-order compact scheme
  for the 3D incompressible Navier-Stokes equations. {\it International Journal
  of Computational Fluid Dynamics} 2017\string; 31(4-5)\string: 214-229.
\newblock \href {\doibase 10.1080/10618562.2017.1326592} {doi:
  10.1080/10618562.2017.1326592}

\bibitem{Moon2020}
Moon H, Hong S, You D. Application of the parallel diagonal dominant algorithm
  for the incompressible Navier-Stokes equations. {\it Journal of Computational
  Physics} 2020\string; 423.
\newblock \href {\doibase 10.1016/j.jcp.2020.109795} {doi:
  10.1016/j.jcp.2020.109795}

\bibitem{LM2015}
Lee M, Moser RD. Direct numerical simulation of turbulent channel flow up to
  {$Re_{\tau} \approx 5200$}. {\it Journal of Fluid Mechanics} 2015\string;
  774\string: 395--415.
\newblock \href {\doibase 10.1017/jfm.2015.268} {doi: 10.1017/jfm.2015.268}

\bibitem{MKM1999}
Moser RD, Kim J, Mansour NN. Direct numerical simulation of turbulent channel
  flow up to {$Re_{\tau}=590$}. {\it Physics of Fluids} 1999\string;
  11(4)\string: 943--945.
\newblock \href {\doibase 10.1063/1.869966} {doi: 10.1063/1.869966}

\bibitem{LJ2014}
Lozano-Dur{\'a}n A, Jim{\'{e}}nez J. Effect of the computational domain on
  direct simulations of turbulent channels up to {$Re_{\tau}=4200$}. {\it
  Physics of Fluids} 2014\string; 26(1)\string: 011702.
\newblock \href {\doibase 10.1063/1.4862918} {doi: 10.1063/1.4862918}

\bibitem{BPO2014}
Bernardini M, Pirozzoli S, Orlandi P. Velocity statistics in turbulent channel
  flow up to {$Re_{\tau}=4000$}. {\it Journal of Fluid Mechanics} 2014\string;
  742\string: 171--191.
\newblock \href {\doibase 10.1017/jfm.2013.674} {doi: 10.1017/jfm.2013.674}

\bibitem{Verzicco2000}
Fadlun EA, Verzicco R, Orlandi P, Mohd-Yusof J. Combined immersed-boundary
  finite-difference methods for three-dimensional complex flow simulations.
  {\it Journal of Computational Physics} 2000\string; 161(1)\string: 35--60.
\newblock \href {\doibase https://doi.org/10.1006/jcph.2000.6484} {doi:
  https://doi.org/10.1006/jcph.2000.6484}

\bibitem{Vreman2014}
Vreman AW, Kuerten JGM. Comparison of direct numerical simulation databases of
  turbulent channel flow at {$Re_{\tau}=180$}. {\it Physics of Fluids}
  2014\string; 26(1)\string: 015102.
\newblock \href {\doibase 10.1063/1.4861064} {doi: 10.1063/1.4861064}

\bibitem{KMM1987}
Kim J, Moin P, Moser R. Turbulence statistics in fully developed channel flow
  at low reynolds number. {\it Journal of Fluid Mechanics} 1987\string;
  177\string: 133--166.
\newblock \href {\doibase 10.1017/S0022112087000892} {doi:
  10.1017/S0022112087000892}

\bibitem{HJ2006}
Hoyas S, Jim{\'{e}}nez J. Scaling of the velocity fluctuations in turbulent
  channels up to {$Re_{\tau}=2003$}. {\it Physics of Fluids} 2006\string;
  18(1)\string: 011702.
\newblock \href {\doibase 10.1063/1.2162185} {doi: 10.1063/1.2162185}

\bibitem{Yamomoto2018}
Yamamoto Y, Tsuji Y. Numerical evidence of logarithmic regions in channel flow
  at {$Re_{\tau}=8000$}. {\it Physical Review Fluids} 2018\string; 3(1)\string:
  012602.
\newblock \href {\doibase 10.1103/PhysRevFluids.3.012602} {doi:
  10.1103/PhysRevFluids.3.012602}

\bibitem{Gholami2016}
Gholami A, Malhotra D, Sundar H, Biros G. FFT, FMM, or Multigrid? A comparative
  study of State-Of-the-Art Poisson solvers for uniform and nonuniform grids in
  the unit cube. {\it SIAM Journal on Scientific Computing} 2016\string;
  38(3)\string: C280--C306.
\newblock \href {\doibase 10.1137/15m1010798} {doi: 10.1137/15m1010798}

\bibitem{Schumann1988}
Schumann U, Sweet RA. Fast Fourier transforms for direct solution of poisson's
  equation with staggered boundary conditions. {\it Journal of Computational
  Physics} 1988\string; 75(1)\string: 123--137.
\newblock \href {\doibase https://doi.org/10.1016/0021-9991(88)90102-7} {doi:
  https://doi.org/10.1016/0021-9991(88)90102-7}

\bibitem{vanderPoel2015}
{van der Poel} EP, Ostilla-M{\'o}nico R, Donners J, Verzicco R. A pencil
  distributed finite difference code for strongly turbulent wall-bounded flows.
  {\it Computers \& Fluids} 2015\string; 116\string: 10--16.
\newblock \href {\doibase https://doi.org/10.1016/j.compfluid.2015.04.007}
  {doi: https://doi.org/10.1016/j.compfluid.2015.04.007}

\bibitem{Zhu2018}
Zhu X, Phillips E, Spandan V, et al. AFiD-GPU: A versatile Navier-Stokes solver
  for wall-bounded turbulent flows on GPU clusters. {\it Computer Physics
  Communications} 2018\string; 229\string: 199--210.
\newblock \href {\doibase https://doi.org/10.1016/j.cpc.2018.03.026} {doi:
  https://doi.org/10.1016/j.cpc.2018.03.026}

\bibitem{Ostilla2016}
Ostilla-M{\'o}nico R, Verzicco R, Grossmann S, Lohse D. The near-wall region of
  highly turbulent Taylor-Couette flow. {\it Journal of Fluid Mechanics}
  2016\string; 788\string: 95--117.
\newblock \href {\doibase 10.1017/jfm.2015.675} {doi: 10.1017/jfm.2015.675}

\bibitem{Costa2018}
Costa P. A FFT-based finite-difference solver for massively-parallel direct
  numerical simulations of turbulent flows. {\it Computers \& Mathematics with
  Applications} 2018\string; 76(8)\string: 1853--1862.
\newblock \href {\doibase https://doi.org/10.1016/j.camwa.2018.07.034} {doi:
  https://doi.org/10.1016/j.camwa.2018.07.034}

\bibitem{PFFT}
Pippig M. PFFT: An extension of FFTW to massively parallel architectures. {\it
  SIAM Journal on Scientific Computing} 2013\string; 35(3)\string: C213--C236.
\newblock \href {\doibase 10.1137/120885887} {doi: 10.1137/120885887}

\bibitem{P3DFFT}
Pekurovsky D. P3DFFT: A framework for parallel computations of Fourier
  transforms in three dimensions. {\it SIAM Journal on Scientific Computing}
  2012\string; 34(4)\string: C192--C209.
\newblock \href {\doibase 10.1137/11082748x} {doi: 10.1137/11082748x}

\bibitem{LiLaizet2010}
Li N, Laizet S. 2DECOMP\&FFT --- A highly scalable 2D decomposition library and
  FFT interface. In: Cray User Group. ; 2010.

\bibitem{Duy2014}
Duy TVT, Ozaki T. A decomposition method with minimum communication amount for
  parallelization of multi-dimensional FFTs. {\it Computer Physics
  Communications} 2014\string; 185(1)\string: 153--164.
\newblock \href {\doibase https://doi.org/10.1016/j.cpc.2013.08.028} {doi:
  https://doi.org/10.1016/j.cpc.2013.08.028}

\bibitem{Song2016}
Song S, Hollingsworth JK. Computation-communication overlap and parameter
  auto-tuning for scalable parallel 3-D FFT. {\it Journal of Computational
  Science} 2016\string; 14\string: 38--50.
\newblock \href {\doibase 10.1016/j.jocs.2015.12.001} {doi:
  10.1016/j.jocs.2015.12.001}

\bibitem{Stone1973}
Stone HS. An Efficient Parallel Algorithm for the Solution of a Tridiagonal
  Linear System of Equations. {\it J. ACM} 1973\string; 20(1)\string: 27--38.
\newblock \href {\doibase 10.1145/321738.321741} {doi: 10.1145/321738.321741}

\bibitem{Hockney1965}
Hockney RW. A fast direct solution of Poisson's equation using Fourier
  analysis. {\it J. ACM} 1965\string; 12(1)\string: 95--113.
\newblock \href {\doibase 10.1145/321250.321259} {doi: 10.1145/321250.321259}

\bibitem{Lawrie1984}
Lawrie DH, Sameh AH. The computation and communication complexity of a parallel
  banded system solver. {\it ACM Trans. Math. Softw.} 1984\string;
  10(2)\string: 185--195.
\newblock \href {\doibase 10.1145/399.401} {doi: 10.1145/399.401}

\bibitem{Wang1981}
Wang HH. A parallel method for tridiagonal equations. {\it ACM Trans. Math.
  Softw.} 1981\string; 7(2)\string: 170--183.
\newblock \href {\doibase 10.1145/355945.355947} {doi: 10.1145/355945.355947}

\bibitem{Sun1989}
Sun XH, Sun HZ, Ni LM. Parallel algorithms for solution of tridiagonal systems
  on multicomputers. In: ICS '89. Association for Computing Machinery.
  Association for Computing Machinery; 1989; New York, NY, USA\string: 303--312

\bibitem{Bao2017}
Bao Y, Luo J, Ye M. Parallel Direct Method of DNS for Two-Dimensional Turbulent
  Rayleigh-Bénard Convection. {\it Journal of Mechanics} 2017\string;
  34(2)\string: 159-166.
\newblock \href {\doibase 10.1017/jmech.2017.54} {doi: 10.1017/jmech.2017.54}

\bibitem{Sherman1950}
Sherman J, Morrison WJ. Adjustment of an inverse matrix corresponding to a
  change in one element of a given matrix. {\it Ann. Math. Statist.}
  1950\string; 21(1)\string: 124--127.
\newblock \href {\doibase 10.1214/aoms/1177729893} {doi:
  10.1214/aoms/1177729893}

\bibitem{Woodbury1950}
Woodbury MA. {\it Inverting modified matrices}.
\newblock Princeton, NJ: Department of Statistics, Princeton University .
\newblock 1950.

\bibitem{Sun1995}
Sun XH. Application and accuracy of the parallel diagonal dominant algorithm.
  {\it Parallel Computing} 1995\string; 21(8)\string: 1241--1267.
\newblock \href {\doibase https://doi.org/10.1016/0167-8191(95)00018-J} {doi:
  https://doi.org/10.1016/0167-8191(95)00018-J}

\bibitem{FFTW3}
Frigo M, Johnson SG. The design and implementation of FFTW3. {\it Proceedings
  of the IEEE} 2005\string; 93(2)\string: 216--231.
\newblock \href {\doibase 10.1109/JPROC.2004.840301} {doi:
  10.1109/JPROC.2004.840301}

\bibitem{IntelMKL}
Intel . Intel Math Kernel Library.
  \url{https://software.intel.com/content/www/us/en/develop/tools/math-kernel-library.html};
  2020.

\bibitem{Top500}
TOP500. \url{https://www.top500.org/lists/top500/2020/11};  2020.

\bibitem{Flores2010}
Flores O, Jim{\'{e}}nez J. Hierarchy of minimal flow units in the logarithmic
  layer. {\it Physics of Fluids} 2010\string; 22(7)\string: 071704.
\newblock \href {\doibase 10.1063/1.3464157} {doi: 10.1063/1.3464157}

\bibitem{Jimenez1991}
Jim{\'{e}}nez J, Moin P. The minimal flow unit in near-wall turbulence. {\it
  Journal of Fluid Mechanics} 1991\string; 225\string: 213--240.
\newblock \href {\doibase 10.1017/S0022112091002033} {doi:
  10.1017/S0022112091002033}

\bibitem{Chen2019}
Chen X, Hussain F, She Z. Non-universal scaling transition of momentum cascade
  in wall turbulence. {\it Journal of Fluid Mechanics} 2019\string; 871.
\newblock \href {\doibase 10.1017/jfm.2019.309} {doi: 10.1017/jfm.2019.309}

\bibitem{Chen2020}
Chen X, Sreenivasan KR. Reynolds number scaling of the peak turbulence
  intensity in wall flows. {\it Journal of Fluid Mechanics} 2020\string; 908.
\newblock \href {\doibase 10.1017/jfm.2020.991} {doi: 10.1017/jfm.2020.991}

\bibitem{Cantwell2019}
Cantwell BJ. A universal velocity profile for smooth wall pipe flow. {\it
  Journal of Fluid Mechanics} 2019\string; 878\string: 834--874.
\newblock \href {\doibase 10.1017/jfm.2019.669} {doi: 10.1017/jfm.2019.669}

\bibitem{Chen2018}
Chen X, Hussain F, She Z. Quantifying wall turbulence via a symmetry approach.
  Part 2. Reynolds stresses. {\it Journal of Fluid Mechanics} 2018\string;
  850\string: 401--438.
\newblock \href {\doibase 10.1017/jfm.2018.405} {doi: 10.1017/jfm.2018.405}

\bibitem{Marusic2010}
Marusic I, McKeon BJ, Monkewitz PA, Nagib HM, Smits AJ, Sreenivasan KR.
  Wall-bounded turbulent flows at high Reynolds numbers: Recent advances and
  key issues. {\it Physics of Fluids} 2010\string; 22(6)\string: 065103.
\newblock \href {\doibase 10.1063/1.3453711} {doi: 10.1063/1.3453711}

\bibitem{Jimenez2012}
Jim{\'{e}}nez J. Cascades in wall-bounded turbulence. {\it Annual Review of
  Fluid Mechanics} 2012\string; 44(1)\string: 27--45.
\newblock \href {\doibase 10.1146/annurev-fluid-120710-101039} {doi:
  10.1146/annurev-fluid-120710-101039}

\bibitem{Yao2018}
Yao J, Chen X, Hussain F. Drag control in wall-bounded turbulent flows via
  spanwise opposed wall-jet forcing. {\it Journal of Fluid Mechanics}
  2018\string; 852\string: 678--709.
\newblock \href {\doibase 10.1017/jfm.2018.553} {doi: 10.1017/jfm.2018.553}

\end{thebibliography}

\clearpage

\end{document}